\documentclass[onecolumn,showpacs]{revtex4}
\usepackage{amssymb}
\usepackage{amsmath,amssymb}

\usepackage{graphicx}

\usepackage{graphics}
\begin{document}
\title{Cosmological Perturbations and the Running Cosmological Constant Model}
\author{Alan M. Velasquez-Toribio
}                     
\address{Departamento de Fisica - ICE, Universidade Federal do Espirito Santo\\
CEP: 29075-910, ES,  Brazil}
\date{Received: date / Revised version: date}
%

\begin{abstract}
We study the matter density fluctuations in the running cosmological
constant (RCC) model using linear perturbations in the longitudinal gauge.
Using this observable, we calculate the growth rate of structures and the matter
power spectrum, and compare these results to $SDSS$ data and
the available data for linear growth rate.
The distribution of collapsed structures may also constrain
models of dark energy. It is shown that the $RCC$ model
enhances departures from the $\Lambda CDM$ model for both cluster
number and cumulative cluster number predicted.
In general, increasing the characteristic parameter $\nu$ leads to
significant growth of the cluster number.
We found that the theory of perturbations provides a useful tool
to distinguish between the new model $RCC$ and the standard cosmological model $\Lambda CDM$.
\end{abstract} 
\maketitle

\section{Introduction}
Recent results from both type Ia supernovae \cite{ries,perl} and the Wilkinson Microwave Anisotropy Probe ($WMAP$), in particular, the five-year data \cite{wmap5,dunk}, indicate that the present Universe is accelerating and that it has negligible spatial curvature. In this context, non-relativistic matter contributes about $30\%$ (dark matter plus baryonic matter) of the critical density of the Universe and the remaining $70\%$ of the energy density is not known and is called dark energy. Dark energy is generally associated with a cosmological constant ($CC$) and can be physically equivalent to the vacuum energy. This component breaks the strong energy condition, but it is the simplest model that one can build. This model, although satisfactory from an observational point of view, is th,eoretically disfavored because there is a huge different between the predicted and observed values of $CC$ \cite{wetal,peeb}. Other possibilities have been considered, among these, the most popular are based on scalar fields, known as quintessence models 
\cite{cal,ford,pad,peebles,ferre,cald,swz,sahni}, and models based on scalar fields with non-canonical kinetic energy, called k-essence models \cite{ar}. Unfortunately, the change in the scalar field may be extremely slow and there is a degenerescence of the potential of quintessence (and k-essence). Another approach is to consider dark energy as an effect of modified gravitation. Some models of this class are, scalar-tensor theory \cite{st}, $F(R)$ theories \cite{fr}, and models that introduce extra-dimensions, such as the $DGP$ model \cite{dgp,lomb}. A recent review of models of dark energy is given in \cite{review,kami}.

In practice, dark energy can be seen into of the Friedmann equations through the relation between its energy density and pressure. This ratio is known as the equation of state ($EOS$) parameter, $w(z)=P(z)/\rho(z)$. The function $w(z)$, where $z$ is the redshift, is a key quantity in attempting to understand the dynamics of cosmological expansion. Another important quantity in the cosmological background is the deceleration parameter q(z). These two are the most commonly used functions in studies of observational constraints using background data. In particular, the deceleration parameter can be used in a model-independent approach (see \cite{daly} and \cite{ishida}).

The next step in the study of a cosmological model is to use linear perturbations. In fact, the behavior of linear perturbations in a scalar field and their effect on large scale structure formations has been investigated by many authors, e.g. see \cite{fejo}. Also the behavior of non–linear gravitational collapse has been investigated \cite{maor,mvan}. These studies are fundamental to understanding and discriminating among competing models.

In this paper, we investigate the cosmological consequences of a model motivated by quantum field theory $QFT$, specifically in the renormalization group. The idea of studying renormalization group effects as a way of solving the $CC$ has been explored previously \cite{Pol81,Pol2001,TV,antmot,lam,jackiw,Reuter2,Reuter3,Reuter4,wetterich,wetterich2,berges,RW}. In particular, a model quadratic for the running of the $CC$ was presented in \cite{fqc,fqc2}, called the running cosmological constant (RCC). Additionally, this model was extended to the logarithmic running of gravitational coupling $G$\cite{Gruni}.
The quadratic model of the running of the $CC$ was solved in reference \cite{CCfit,CCfit2,CCfit3}. In this model, the energy density of the vacuum $\rho_{\Lambda}(z)$ can be given as a quadratic function of the expansion rate, which allows one to unambiguously define the equation of conservation of energy and determine the hubble parameter $H(z)$ as function of the redshift. Reference \cite{ss} generalizes this model to run for $CC$ and $G$. A possible fundamental relation with $QFT$ was first proposed in reference \cite{sola1} for running $CC$ and a logarithmic run of the gravitational coupling. 
Let us also mention the reference \cite{mm} where the quadratic evolution law of the $CC$ is also emphasized from
the point of view of $QFT$ in curved space-time using the effect of zero-point fluctuations in a FLRW background. This model is confronted with the latest accurate observational data in \cite{me}. 
Additionaly, it is worthwhile to note that the paper \cite{nb} reaches similar
conclusions for the evolution of the CC from the point of view of
supersymmetric theories.

The model of the running of the $CC$ is equivalent to models of dark energy in the fluid approximation (parameterizations of the $EOS$ and quintessence models \cite{ilso}). The background cosmology of this model has been well investigated with data from type Ia supernovae, restricting the values of parameter $\nu$, which represents the "run" of the $CC$ in the $RCC$ model. A recent review of the class of these models within QFT in curved space-time is given in \cite{jsola} and \cite{ilyas}. Additionally, reference \cite{fabris} determined the matter power spectrum in the synchronous gauge.

We investigate matter density perturbations using the longitudinal gauge. This approach was pioneered in \cite{mukanov,MFB}. Their formalism is applied to our $RCC$ model to determine the linear growth rate and the matter power spectrum as functions of parameter $\nu$.  We compared the results to observational data from the SDSS (Sloan Digital Sky Survey). Recently, the running of the $CC$ and of gravitational coupling using the longitudinal gauge was studied \cite{tp}. Some comments between this and our work are given in the conclusion section.

On the other hand, it has been recognized in various papers and simulations \cite{evrard,evrard2,evrard3,evrard4,evrard5} that the evolution of the cluster number counts can determine the properties of dark energy. Therefore, we use our results for linear density perturbations and the Press-Schechter formalism to determine the mass function, the number counts, and the cumulative number counts as functions of redshift, and study the sensitivity of parameter $\nu$. We use the $\Lambda CDM$ model to compare our predictions. Additionally, we investigate how the number counts and cumulative number counts depends on the parameters $\Omega_{m0}$ and $h$ when the value of $\nu$ is fixed. Similar quantities were studied in \cite{rcg}.
Some considerations will be made in the text.

Our paper is organized as follows. In section II, we introduce the $RCC$
model and discuss the behavior of the comoving energy density and
deceleration parameter. In section III, we discuss the linear
perturbations of matter. Section IV is devoted to the computation of cluster number counts in the Press-Schechter formalism. In section V, we present
our conclusions. In the appendix, we display results for the $DGP$ model
and scalar-tensor theory, both of which are useful when calculating the linear
growth rate.

\section{Running Cosmological Constant}
\label{sec:1}

In this section, we will introduce our cosmological model by
considering a $FLRW$ metric and a Universe of
matter (baryonic + dark) and dark energy or
vacuum energy; thus, the cosmological evolution is governed by the
following Friedmann equation:
\begin{equation}
\label{eq1}
H^{2} = \displaystyle \frac{8 \pi G}{3}\left(\rho_{m}+ \rho_{\Lambda}\right) - \frac{k}{a^{2}},
\end{equation}
where $\rho_{m}$ and $\rho_{\Lambda}$ are the densities of matter and energy,
respectively, $a$ is the scale factor, $H=\frac{\dot{a}}{a}$
is the Hubble parameter, and $k$ is a constant that can take the values $+1,0,-1$. In this investigation, we restrict ourselves to an evaluation of
the flat case. Using the Bianchi identity, we can write the conservation law:
\begin{equation}
\label{eq2}
\frac{d \rho_{m}}{dz} + \frac{d \rho_{\Lambda}}{dz} = 3\frac{\rho_{m}}{1+z}.
\end{equation}

In the above equation, we can see that there is an exchange of energy
between the two components of the Universe, so both are completely linked. The dark
matter can be subject to an energy exchange
resulting in both a time-dependent mass and a modification of its equation of state.
Therefore, in this model, the dynamic of the Universe is
dominated both by the evolution of the $CC$ and by its interaction with
dark matter.

On the other hand, an important issue to be noted is that equations (\ref{eq1}-\ref{eq2})
do not form a complete set of equations because
there are three free variables, $\rho_{m}(z), \rho_{\Lambda}(z), H(z)$ and
two equations. We need a third equation to have a complete system.
In this paper, we introduce an additional equation by considering the effects of
applying the renormalization group to the $CC$.
In this framework, the energy density in the $RCC$ model is \cite{fqc,CCfit,Gruni}:
\begin{equation}
\label{eq3}
\frac{d \rho_{\Lambda}}{d \ln H} = \frac{\sigma H^{2} M^{2}}{(4 \pi)^{2}}.
\end{equation}
The above equation was proposed based on the assumption that the renormalization group scale $\mu$ is identified
with $H(z)$. This scale was originally proposed in \cite{fqc} and can be considered as purely phenomenological ansatzs. $M$ is an effective mass parameter representing the average mass of the heavy particles in the grand unified
theory (GUT) near the Planck scale, after taking into account their multiplicities.
The coefficient $\sigma$ can be positive or negative, the sign depends on whether bosons ($\sigma = +1$) or fermions ($\sigma = -1$) dominate in the loop contributon,
this is, it depends on whether fermions
or bosons dominate at the highest energies. Recall that, in the renormalization group framework, equation (\ref{eq3})
is interpreted as a
"$\beta$-function" of $QFT$ in curved space-time, and that it determines the running of the $CC$, for other details see \cite{basi}. Therefore, in the background (cosmology of zeroth order), the three
equations above allow a full description of the evolution of the $RCC$
cosmological model.
Using equation (\ref{eq3}), we can obtain one explicit solution for $\rho_{\Lambda}$:
\begin{equation}
\label{eq4}
\rho_{\Lambda}=\rho_{\Lambda 0}+\frac{3\nu}{8 \pi}M_{p}^{2}\left(H^{2}-H^{2}_{0}\right),
\end{equation}
where $\rho_{\Lambda 0}$ and $H_{0}$ are the current values of these parameters.
Additionally, in this model, we find a new parameter $\nu$ (dimensionless)
that is given by
\begin{equation}
\label{eq5}
\nu \equiv \frac{\sigma}{12 \pi}\frac{M^{2}}{M_{p}^{2}},
\end{equation}
thus, from equation (\ref{eq4}), if $\nu$ is zero, the effects
of a run are canceled and we recover the standard model, $\Lambda CDM$.

\begin{figure}
\begin{center}
\includegraphics[height= 6.8 cm,width= 9.0cm]{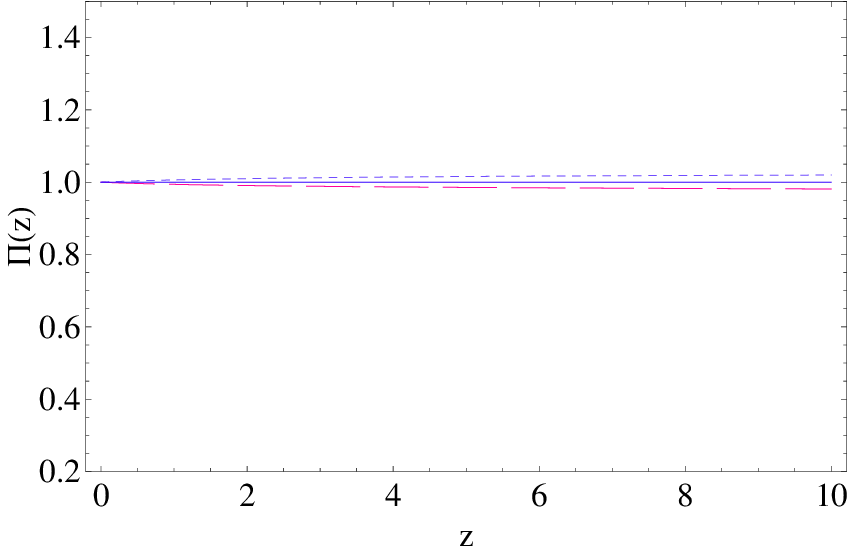}
\end{center}
\caption{Comoving background matter density as a function of redshift in the
RCC model. We can see that the decrease in density is an indicator
of the coupling between dark matter and vacuum energy. Note that, in this
plot, $\Lambda CDM$ corresponds to a constant line equal to one. From bottom to top: $\nu=0.25$ (orange), $=0.065$ (red), $=10^{-3}$ (blue).
In all models we used $\Omega_{m0}=0.24$ and $h=0.72$.}
\label{fig1}
\end{figure}

\begin{figure}[htb]
\begin{center}
\includegraphics[height= 6.8 cm,width= 10.0cm]{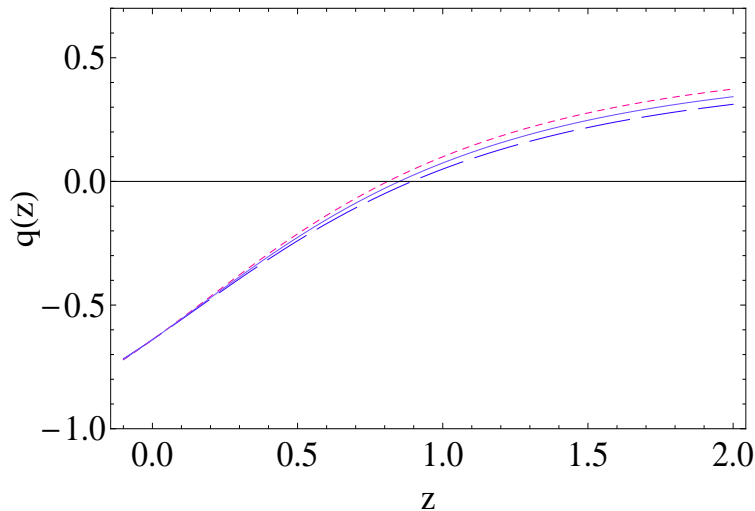}
\end{center}
\caption{The plot of deceleration parameter $q(z)$ in the $RCC$ model. We can see that the deceleration parameter is slightly
sensitive to the change of sign of the parameter.
We used $\Omega_{m0}=0.24$ and $h=0.72$. }
\label{fig2}
\end{figure}

Based on the above equations, equation (\ref{eq1}) can be rewritten as
\begin{eqnarray}
\label{eq7}
\frac{H^{2}}{H^{2}_{0}} &=& 1 + \left(\Omega_{m0}-\frac{2\nu\Omega_{k0}}{1-3\nu}\right)\left(\frac{(1+z)^{3-3\nu} -1}{1-\nu}\right)+
\frac{\Omega_{k0}(z^2 + 2z)}{1-3\nu},
\end{eqnarray}
where $H_{0}^{2}\Omega_{k0}(1+z)^{2}=-\frac{k}{a^{2}}$ and $\Omega_{m0}=\Omega_{MD}+\Omega_{B0}$, where
$\Omega_{DM}$ is the parameter for dark matter and $\Omega_{B0}$ is the parameter for baryonic matter.
This expression for the Hubble parameter is valid for Universes with positive
or negative curvature. Therefore, equations (\ref{eq1}), (\ref{eq4}), and (\ref{eq7})
define our cosmological model. One of the first quantities that we can
calculate is the matter density. We define the comoving matter
density function as $\Pi(z) = \frac{\rho_{m}(z)}{(1+z)^{3}}$. In Figure 1,
we displayed it. There is a decrease in the density caused by the coupling
between dark matter and dark energy. Increasing the coupling leads to a faster
decrease in the density. In Figure 1, the constant line corresponds to $\Lambda CDM$
($\nu=0$). Note that this behavior is similar to that of
coupled quintessence models \cite{mota}. We used only positive values for parameter $\nu$.
For negative values, the density $ \Pi $ is greater than one.

Another important parameter in the cosmology of the background is the deceleration
parameter $q(z)=-1-H(1+z)\frac{d}{dz}(\frac{1}{H})$. In Figure 2, we show this parameter
for our $RCC$ model. When $z>1$, in all cases (negative or positive) the deceleration parameter is
positive and tends rapidly toward a matter dominated phase. For values close to $z = 0$, the deceleration parameter can not distinguish between positive and negative values of $\nu$. In the next section, we consider
linear perturbations and structure formation.

\section{The Linear Perturbation Equations}
The theory of cosmological perturbations is based on the expansion of Einstein's equations to linear order around the background metric. The decomposition theorem states that perturbations in the metric can be divided into three type: scalar, vector, and tensor. In this paper, we will consider scalar perturbations, specifically, density perturbations in the $RCC$ model because we are interested in the matter power spectrum. Another fundamental question is the choice of gauge or coordinate system. General Relativity leads to the question of gauge freedom. This means that if we change the coordinate system we use, we would get a metric of a different form. One way of dealing with the gauge problem is to eliminate gauge dependence entirely. This approach is referred to as using gauge-invariant variables and was pioneered by Bardeen \cite{bardeen}. However, in the literature, many other gauges have been used \cite{weinberg}. For example, the dynamics of density perturbations in the $RCC$ model were investigated in the synchronous gauge \cite{fabris}. In the present work, we are extending the analysis of \cite{fabris} by computing the matter power spectrum in the longitudinal gauge and we will study the cluster number counts.

The longitudinal gauge is, from the physical point of view, much more intuitive because metric perturbations are similar to Newtonian perturbations. This gauge is commonly chosen for work in CMB and gravitational lensing. On the other hand, conceptually, this gauge fixes all spurious degrees of freedom, and the two scalar potentials $\Psi$ and $\Phi$ that appear in the line element correspond to Bardeen's gauge invariants \cite{MFB,bardeen}; this is, however, not the case of the synchronous gauge where there exist residual transformations that lead to the appearance of unphysical solutions. However, the use of this gauge is justified because these spurious modes are canceled to some extent when calculating a physical observable, which, by definition, cannot depend on a given system of coordinates. The metric in the longitudinal gauge is given by \cite{mukanov}:

\begin{equation}
\label{eq8}
ds^{2} =  -(1+2\Psi(\vec{x},t))dt^2 + (1+2\Phi(\vec{x},t))dx_{j}dx^{j}.
\end{equation}

It should be noted that the longitudinal gauge is restricted to scalar modes; nonetheless, it can be easily generalized to include the vector and tensor degrees of freedom \cite{bertschinger}. Further, in the absence of anisotropic stress, one of Einstein's equations gives $\Psi = -\Phi$; the two gravitational potentials are equal and opposite \cite{dodelson}. Therefore, there remains only one free metric perturbation variable, which is a generalization of the Newtonian gravitational potential. This justifies the name of Newtonian gauge.

To derive the equations for the density perturbations, we follow the standard formalism \cite{bertschinger,weinberg,dodelson}. We consider the entropy perturbation to be negligible and the energy-momentum tensor to be free of anisotropic stresses. Thus, in these conditions, the energy-momentum tensor has the form of a perfect fluid,
\begin{eqnarray}
\label{eq9}
T^{\mu}_{\nu}&=&\displaystyle pg^{\mu}_{\nu}+\left(\rho+p\right)U^{\mu}U_{\nu},
\end{eqnarray}
where  $U^{\mu}=dx^{\mu}/(-ds^2)^{1/2}$ is the four-velocity of fluid, $p$ is the pressure, and $\rho$ is the energy density of a perfect fluid. For a fluid moving with a small velocity $v^{i} \equiv dx^{i}/d\tau$ (peculiar velocity), the $v^{i}$ can be treated as a perturbation of the same order either as $\delta \rho = \rho - \bar{\rho}$ or as $\delta p = p - \bar{p}$. The quantities $\bar{\rho}$ and $\bar{p}$ refer to the background. In linear order, the perturbations of the energy-momentum tensor that we used are given by:
\begin{eqnarray}
\label{eq10}
T^{0}_{0}&=&-\displaystyle(\bar{\rho}+\delta \rho \displaystyle), \nonumber \\
T^{0}_{i} &=& \displaystyle(\bar{\rho}+\bar{p}\displaystyle)v_{i} = -T^{i}_{0}, \\
T^{i}_{j} &=& \displaystyle(\bar{p}+\delta \bar{p}\displaystyle)\delta^{i}_{j}. \nonumber
\end{eqnarray}
The perturbed four-velocity is
\begin{eqnarray}
\label{eq11}
U^{\alpha} &=& \displaystyle((1-\Psi), v^{i}\displaystyle).
\end{eqnarray}

Although one can directly work with the Einstein's equations, it turns out to be convenient to use the equations of motion for the matter variables because we are ultimately interested in matter perturbations. We consider the conservation of the energy-momentum tensor as
\begin{eqnarray}
\label{eq12}
T^{\alpha}_{\beta;\alpha}=\frac{\partial T^{\alpha}_{\beta}}{\partial x^{\alpha}} + \Gamma^{\alpha}_{\delta \alpha}
T^{\delta}_{\beta}-\Gamma^{\gamma}_{\beta \alpha}T^{\alpha}_{\delta}=0.
\end{eqnarray}
This expression gives us two equations, one for $\beta = 0$ and the other for $\beta = i$. To determine these equations, we consider the perturbations for the two densities and the metric
\begin{eqnarray}
\label{eq13}
\rho_{m} &\rightarrow& \rho_{m}\left(1+\delta_{m}\right), \nonumber \\
\rho_{\Lambda} &\rightarrow& \rho_{\Lambda}(1+\delta_{\Lambda}),\\
g_{\mu \nu} &\rightarrow& g_{\mu \nu} + h_{\mu \nu}. \nonumber
\end{eqnarray}

For the sake of simplicity, we want to make full use of the symmetry under spatial translations; this can best be exploited by working with the Fourier components of perturbations. Therefore, we have
\begin{eqnarray}
\label{eq14}
f(\vec{x}) &=& \int{\frac{d^{3}k}{(2\pi)^3}e^{i\vec{k}.\vec{x}}F(\vec{k})}.
\end{eqnarray}
With these considerations, we use the line element (equation \ref{eq8}), equations (\ref{eq10}) and (\ref{eq13}) put into equation (\ref{eq12}), and we get
\begin{eqnarray}
\label{eq15}
\dot{\delta\rho_{m}}+ \dot{\delta\rho_{\Lambda}} + \rho_{m}\left(\theta+3\Psi_{,0}\right) + 3H\delta\rho_{m}&=&0,\\
\dot{\rho_{m}} \theta + \rho_{m} \dot{\theta}- \frac{k^{2}\Psi}{2} + 3H\rho_{m}\theta &=& -(\frac{k}{a})^{2} \delta\rho_{\Lambda},
\end{eqnarray}
where the dot is the derivative with respect to cosmic time and $\theta=\partial_{i} v_{i}$. The energy density in the $RCC$ model, $\rho_{\Lambda}$, can be written as \cite{fabris}
\begin{eqnarray}
\label{eq16}
\rho_{\Lambda}&=&A + B \displaystyle(\nabla_{\mu}U^{\mu}\displaystyle)^2,
\end{eqnarray}
where we have used the fact that $\nabla_{\mu}U^{\mu}=3H$ and defined $A=\rho_{\Lambda 0} - \frac{3\nu}{8\pi}M_{p}^{2}H_{0}^{2}$ and $B=\frac{\nu M_{p}^{2}}{24 \pi}$. It is important to note that the expression above is not unique; although, it has a compact form and is useful for investigating linear perturbations. The forms of the terms $A$ and $B$ have specific forms as functions of the $\nu$ parameter. Additionally, the form of (16) is the simplest quadratic form, that is, a constant plus constant by quadratic form. It is important to note that other papers have used different methods for constructing the perturbation equations (see \cite{rcg},\cite{pelinson1}).
Using the perturbed four-velocity, equation (\ref{eq11}), the perturbed Christoffel symbols, and keeping only the linear order term, we find
\begin{eqnarray}
\label{eq17}
\delta_{\Lambda}  &=& \frac{2H\nu \displaystyle[\theta - 3(\dot{\Psi}+ H\Psi)\displaystyle]}{\rho_{\Lambda}(z)}.
\end{eqnarray}
By contrast, the Einstein equations in the longitudinal gauge are
\begin{eqnarray}
\label{eq18}
k^{2}\Psi + 3H(z)(\dot{\Psi}+ H(z) \Psi) &=& -4\pi G a^{2} (\delta_{m} \rho_{m}(z) + \delta_{\Lambda} \rho_{\Lambda}(z)),\\
\dot{\Psi}+ H(z)\Psi &=& -4\pi G a \rho_{m}\theta.
\end{eqnarray}
Substituting equation (20) into equation (19), we obtain a new equation for the scalar potential as a function of the variables of matter
\begin{eqnarray}
\label{eq19}
k^{2}\Psi &=& \frac{3H\rho_{m}\theta}{2k^{2}(1+z)^{3}}- \frac{(\delta_{m}\rho_{m}+ \delta_{\Lambda} \rho_{\Lambda})}{2(1+z)^{2}}.
\end{eqnarray}

It is convenient for our numerical calculations to write equations (15), (16), (18), and (21) in terms of the redshift, i.e., using $dt=\frac{dz}{H(1+z)}$. Further, it is advantageous to use the following two ratios \cite{fabris},
\begin{eqnarray}
f_{1}(z) &=& \frac{\rho_{m}}{\rho_{t}} = \frac{(1+z)2H(H')-
2H_{0}^{2}\Omega_{k0}(1+z)^{2}}{3(H^{2}-H_{0}^{2}\Omega_{k0}(1+z)^{2})}, \nonumber \\
\\
f_{2}(z) &=& \frac{\rho_{\Lambda}}{\rho_{t}} = \frac{3H^{2}-2H(1+z)H'-H_{0}^{2}\Omega_{k0}(1+z)^{2}}{3(H^{2}-H_{0}^{2}\Omega_{k0}(1+z)^{2})}, \nonumber
\end{eqnarray}
and
\begin{eqnarray}
\dot{\Psi} -H\Psi &=& \frac{\rho_{t} v}{2k^{2}(1+z)},
\end{eqnarray}
\begin{eqnarray}
\theta -3\dot{\Psi}&=&\frac{v}{f_{1}}+ 3H\left(\Psi + \frac{\rho_{t} v}{2(1+z) H k^{2}}\right),
\end{eqnarray}
where $\rho_{t}(z) = 3H^{2}(z) - 3H_{0}^{2}\Omega_{k0}(1+z)^{2}$. Using equations (22-24), the system of equations for the perturbations can be written as

\begin{subequations}
\label{allsys}
\begin{eqnarray}
&&\delta_{\Lambda}(z) = \frac{\nu v H}{\rho_{t}f_{2}}\displaystyle[\frac{1}{f_{1}}- \frac{3\rho_{t}}{2k^{2}(1+z)}\displaystyle]\,,\\
\nonumber \\
&&\delta_{m}^{'}(z)+\delta_{m}(z)\left[\frac{f^{'}_{1}}{f_{1}}+\frac{3f_{1}}{1+z}-\frac{3}{(1+z)}\right]+
\frac{3f_{2}}{(1+z)}\delta_{\Lambda}+\frac{3\Psi}{(1+z)}+\frac{v}{H(1+z)}\left(\frac{1}{f_{1}}+\frac{3\rho_{t}}{k^{2}(1+z)}\right) \nonumber \\
&& +\frac{2\nu}{f_{1}}\left(v f'_{2}+f_{2}v'\right)(K(z)+\frac{M(z)}{2})+
\frac{2\nu f_{2}v}{f_{1}}\left(K'(z) + \frac{M'(z)}{2}\right)=0\,, \nonumber \\
\nonumber \\
&&v{'}(z)+\frac{3(f_{1}-1)v}{1+z} = \frac{k^{2}(1+z)}{H}\left(\delta_{\Lambda}f_{2} - \frac{\Psi}{9H^{2}}\right) \,,\\
\nonumber \\
&&k^{2}\Psi=\frac{\rho_{t}}{2(1+z)^{2}}\left[\frac{3v}{k^{2}(1+z)}-\left(\delta_{m}(z)f_{1}+
 \delta_{\Lambda}(z)f_{2}\right)\right]\,,
\end{eqnarray}
\end{subequations}

where the prime refers to the derivative with respect to redshift, $v= f_{1}\theta$, and the $M(z)$ and $K(z)$ terms of the equation (\ref{allsys}b) are given by
\begin{eqnarray}
M(z) &=& \frac{1}{f_{2}k^{2}(1+z)},
\end{eqnarray}
\begin{eqnarray}
K(z) &=& \frac{1}{3Hf_{2}f_{1}}.
\end{eqnarray}
In these equations, $\nu$ is the parameter defined in (5) and is very important because when $\nu=0$, the perturbation in the vacuum energy is canceled. In this way, we recover the $\Lambda CDM$ scenario as a particular case. In equation (24), is important to consider some aspects of gauge dependence. In reference \cite{fabris}, the linear perturbation equations in the synchronous gauge were determined, equations in which there is no scale dependence explicitly in the equation for $\delta_{m}$ (see equation (3.19) of reference \cite{fabris}). On the other hand, in equation (24), there are terms that explicitly contain $k$; however, the two gauges clearly agree at small scales, where $1/k \rightarrow 0$. At large scales, these equations are not equivalent. The Newtonian gauge includes the expansion of the Universe and is, therefore, more appropriate for a description of large scale perturbations because it corresponds to a time slicing of isotropic expansion \cite{MFB}. The synchronous gauge corresponds to a time slicing obtained for a free falling observer frame. Additionally, as mentioned at the beginning of the section, the Newtonian gauge is directly related to the gauge invariant quantities in the approximation of Bardeen \cite{bardeen}. Therefore, the Newtonian gauge is more important from the observational point of view; however, strictly speaking, it is necessary to make some changes when we intend to use the theoretical quantity $\delta_{m}$ together with observations. Recently, in reference \cite{yz},  the matter power spectrum was investigated using gauge-dependent quantities which could introduce gauge modes that can cause artificial large scale enhancement of the power spectrum. For an accurate determination of the matter power spectrum, it is essential to consider the effects of redshift space distortions (caused by the peculiar velocities), the scale dependence of the galaxy bias, and magnification by gravitational lensing \cite{ayz,ayz2,ayz3}. In the following sections we used the theoretical quantity
$\delta_{m}$ in the Newtonian gauge and $SDSS$ data for the matter power spectrum, ignoring the above aspects.

\begin{figure*}[!h]
\begin{center}
\includegraphics[height= 7. cm,width= 8.5cm]{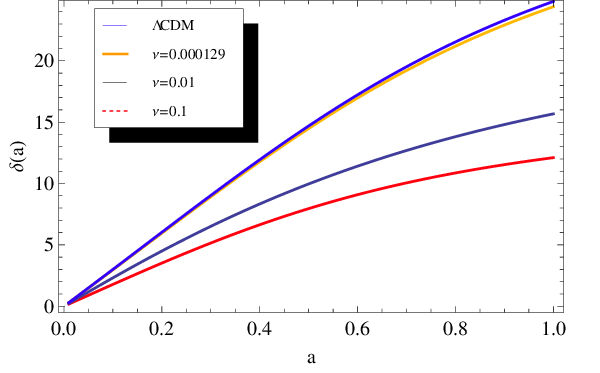}
\includegraphics[height= 9. cm,width= 8.5cm]{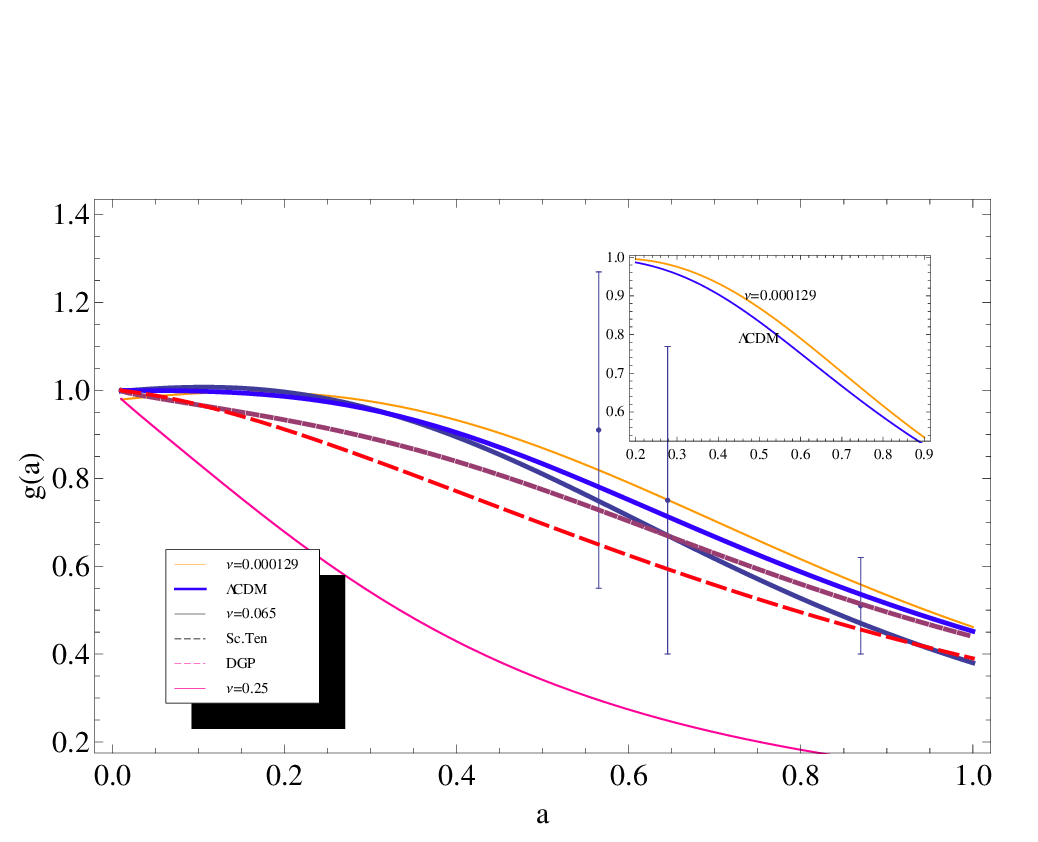}
\end{center}
\caption{
The growth factor (left) and the linear growth rate (right) for the $RCC$ model. We also show the linear growth rate for the $DGP$ and scalar-tensor models (see appendix for details). In all cases, we use $\Omega_{m0}=0.24$ and $h=0.72$.}
\label{fig3}
\end{figure*}

\subsection{The Linear Growth Rate}

The solution to system (25) allows us to determine the density contrast of matter, $\delta_{m}$, which is necessary to determine the growth factor, defined as
\begin{equation}
D(a) = \frac{\delta_{m}(a)}{\delta_{m}(a=1)}.
\end{equation}

In Figure 3 (left) we show the growth factor in the $RCC$ model for different values of $\nu$; we can see that for $\nu=10^{-4}$ there is concordance with a $\Lambda CDM$ model. However, observationally, it is more important to know the linear growth rate, which measures how rapidly structure is being assembled in the Universe as a function of cosmic time (scale factor or redshift). Linear growth rate is defined by
\begin{equation}
g(a)=\frac{d\ln D(a)}{d\ln a}.
\end{equation}

This quantity has been measured using different catalogs. In general, the redshift maps of galaxies are distorted by the peculiar velocities of galaxies along the line of sight. At large scales, this distortion can be expressed through the redshift distortion parameter $\beta$ and can be shown to be related to the linear growth rate as \cite{hamilton}
\begin{equation}
\beta = \frac{g(z)}{b_{L}},
\end{equation}
where $b_{L}$ is the linear bias value ($b_{L} = \sigma_{R}^{gal}/\sigma_{R}^{mass}$), that is, the ratio between the root-mean-squared ($rms$) density contrasts in the galaxy and mass distributions on scale $R$ where linear theory applies. Therefore, a measure of $g(z)$ can be obtained using these two parameters. The $\beta$ parameter may be measured from redshift surveys by measuring the power spectrum of the galaxies \cite {hamilton} and the $b_{L}$ can be obtained from the skewness induced in the bispectrum of a given survey. Using this technique, values of $\beta$ and $b_{L}$ have been measured using the 2dFGRS sample of 220,000 galaxies \cite{verde}. Recently, another measure was made using the spectroscopic data from the VIMOS-LT Deep Survey \cite{nature}. Finally, another measure of growth rate was made using the 2dF-SDSS LRG and QSO survey \cite{ross}.
However, in this last case, the value of $\beta$ and $b_{L}$ are not fully independent, because they have been obtained by imposing simultaneous consistency with the clustering measured at $z=0$. All these data are compiled in table 1. In Figure 3 (right) we show these estimates of the growth rate and compare them to predictions from various theoretical models. We plot the linear growth rate for the $DGP$ and scalar-tensor models. Here, we use the results shown in the appendix. Despite the large error bars, the measurements indicate the need for a small value for parameter $\nu$, and hence, are very close to the $\Lambda CDM$ model.

\begin{table*}
\begin{center}
\caption{\label{tab:table2}In this table we compile the different values of the linear
growth rate found in the literature.}

\begin{tabular}{ccccc}
 Survey&$\beta \pm \delta \beta$ &$b_{L} \pm \delta b_{L}$&$ g \pm \delta g$& z\\
\hline
2dFGRS \cite{verde}& $0.49 \pm 0.09$ & $1.04 \pm 0.10$ & $0.49 \pm 0.14$ & 0.15
 \\
2dF-SDSS LRG and QSO \cite{ross}& $0.45 \pm 0.05$ & $1.66 \pm 0.35$ & $0.75 \pm 0.35$ &0.55
 \\
  VVDS \cite{nature}& $0.70 \pm 0.26$ & $1.30 \pm 0.10$ &  $0.91  \pm 0.36$  &0.80
\\
\end{tabular}
\end{center}
\end{table*}

\subsection{The Matter Power Spectrum}

Another important amount that we can determine from the set of equations in (25) is the matter power spectrum, defined as
\begin{equation}
P(k,z) = \delta^{2}(k,z).
\end{equation}
To set the initial conditions, we can use the $BBKS$ approximation for the transfer function \cite{bbks},
\begin{equation}
T(k) = \frac{\ln(1+2.34q(k))}{(2.34q(k))[1+3.89q+(16.1q)^{2}+(5.46q)^{3}+(6.71q)^{4}]^{1/4}},
\end{equation}

in the presence of $CC$, we know that $q=k/(\Sigma h Mpc^{-1})$, where the shape parameter is $\Sigma = \Omega_{m0}he^{-\Omega_{b0}}$ and $h = H_{0}/100$ because observable wavenumbers are in units of $hMpc^{-1}$. Further, we assume that only $4\%$ of the cosmic density is provided by conventional baryonic matter. The matter power spectrum can be written in the form
\begin{equation}
P(k)=Ak^{n}T^{2}(k),
\end{equation}
where $n$ measures the slope of the primordial power spectrum (we will assume $n=1$ \cite{wmap5,sl}), and $A$ is a normalization constant. To obtain power spectra for our $RCC$ model, we have performed the numerical analysis using the equations in (25) from $z = 500$ to $z = 0$. At $z=500$, matter is dominant. In normalizing the power spectrum, we followed the methodology presented in \cite{fabris}.

In Figure 4, we present the matter power spectrum for our model, where $\Omega_{m0}=0.24$ and we use the power spectrum estimated by Percival et al. based on data from the SDSS Project. The impact that the $RCC$ model has on the linear power spectrum is dominated by equation (24a) and depends on the value of the parameter $\nu$. If parameter $\nu$ is zero, we recover the $\Lambda CDM$ model, because $\delta_{\Lambda} = 0$. Consequently, a large value of $\nu$ entails a greater damping of the power spectrum in the case of $\nu$ being positive. In the case of $\nu$ being negative, the deviation of the power spectrum with respect to the observational data is even stronger. In principle, this feature can also be seen in Figure 4. We can interpret this behavior theoretically because in equation (15), the parameter $\nu$ appears as a factor in the term $k^{2}$ is proportional which is derived from the pressure gradient. When $\nu$ is negative, this term changes sign. Thus, it is expected that for small scales ($k> 0.15$) and $\nu<0$, the pressure terms become important. In general, if we consider large scales ($k <0.06$, for example) the $k^{2}$ terms may be negligible.

\begin{figure*}[htb]
\begin{center}
\includegraphics[height= 9.5 cm,width=12.5cm]{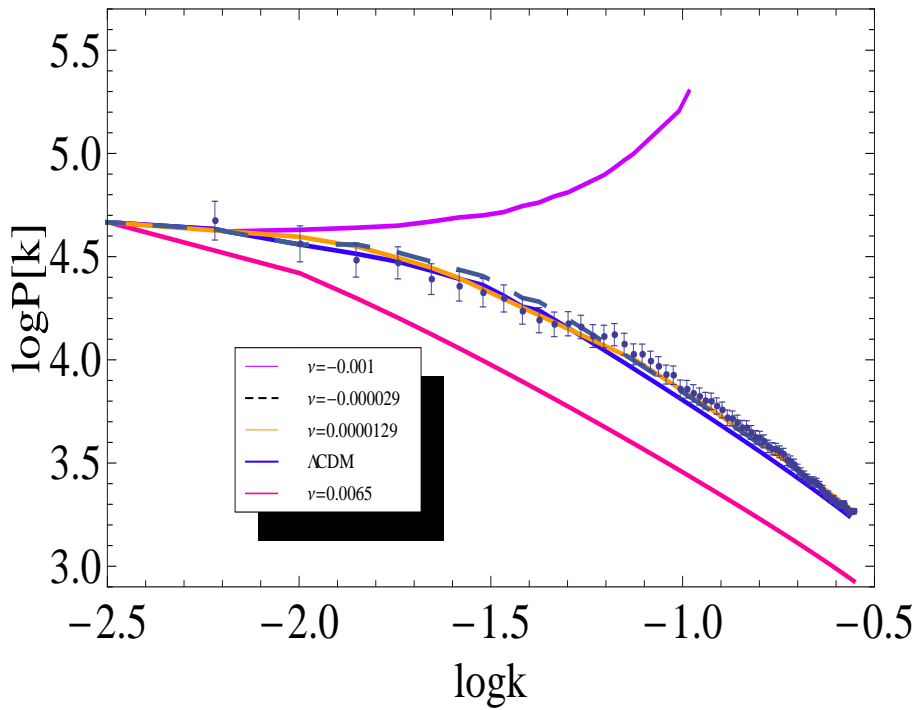}
\end{center}
\caption{The matter power spectrum for the $RCC$ model. The blue curve is the $\Lambda CDM$ and is very close to the $RCC$ model with the best fit $\nu=(1.29 \pm 2.39) \times 10^{-5}$.  We used $\Omega_{m0}=0.24$ and $h=0.72$.}
\label{fig4}
\end{figure*}

We can determine the mass in terms of the Planck mass for $\nu \approx 10^{-5}$ as $M \approx 0.6 \times 10 ^{-2} M_ {p}$, that is, approximately to the GUT particle spectrum. Therefore, the $RCC$ model is viable for masses smaller than the Planck mass.

To get a more accurate value of $\nu$ (without using the rigorous theory of parameter estimation), we use the $\chi^{2}$ statistic. We determined the quality of the fit between the theoretical estimate and the observational data, which is defined as: $\chi^{2}=\Sigma_{i}{\frac{(P_{ob}-P_{the})^{2}}{\sigma_{ob}^{2}}}$, where
$P_{ob}$ is the observational value of the power spectrum, $P_{the}$ is the corresponding theoretical result, and $\sigma_{ob}$ denotes the observational error bars. Therefore, the probability distribution function can be defined as $P(\nu)= F_{0}e^{-\chi^{2}/2}$, where $F_{0}$ is a normalization constant. Minimizing $\chi^{2}$, we can determine the most probable value of parameter $\nu$. In our case, this value is $\nu=(1.29 \pm 2.39) \times 10^{-5}$, where we have assumed that $\Omega_{m0}=0.24$ and $h=0.72$.

In the next section, we consider only the case of $\nu >0$, and a detailed analysis of the case of $\nu<0$ will be presented in the future. In this paper our aim is to understand the overall behavior of the $RCC$ model, rather than determine the observational constraints, which requires a rigorous application of the theory of parameter estimation \cite{press1}.

\section{The Number Counts}

It has long been recognized that modeling a cluster of galaxies provides a useful test of
the fundamental cosmological parameters. The total abundance of cluster $N$ and its
distribution in redshift $\frac{dN}{dz}$ should be determined by the geometry of the
Universe and the power spectrum of initial density fluctuations. One of the first
cosmological parameters to be constrained was $\sigma_{8}$, the amplitude of
mass density fluctuations on a scale of 8 $h^{-1} Mpc$. For example, recently, Komatsu et al. determined a value of $\sigma_{8} = 0.812\pm0.026$ using the data from WMAP. However, in general, the value of $\sigma_{8}$ is inaccurate (for example, there is an implicit uncertainty in the value of $\sigma_{8}$ as a function of $w$; see \cite{kt,kt2}).

Our objective in the present section is to determine the clusters number count and its
evolution with redshift and investigate whether these quantities depend
significantly on $\nu$. We used the
Press-Schechter ($PS$) \cite{ps,bond} formalism to give a prescription for
estimating the mass function for a hierarchical gaussian density field.

\begin{figure*}[!h]
\begin{center}
\includegraphics[height= 7.0 cm,width=7.5cm]{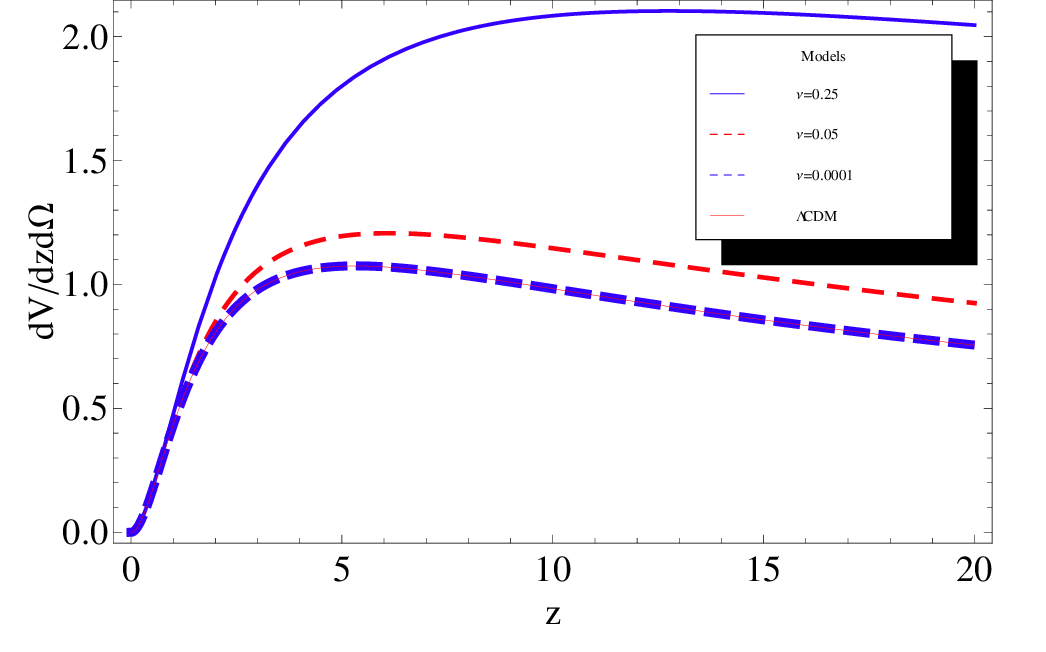}
\includegraphics[height= 7.0 cm,width=7.5cm]{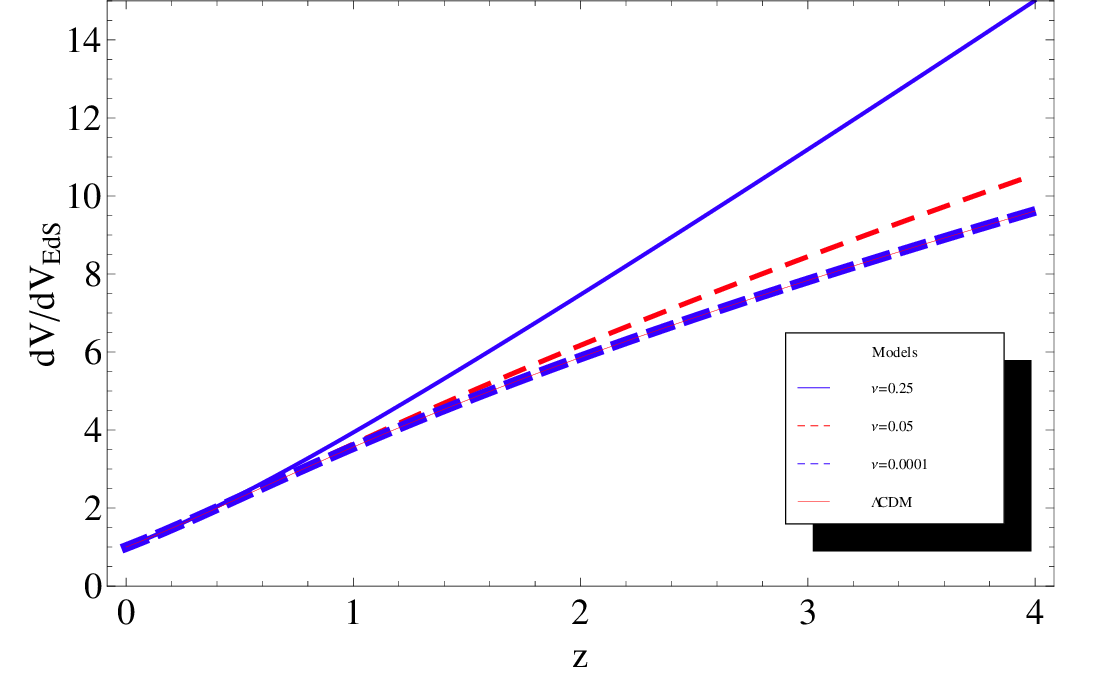}
\end{center}
\caption{The figure shows the evolution of the comoving volume element with
redshift for values different of $\nu$.
We show (below) the volume element with respect to the de-Sitter volume.}
\label{fig5}
\end{figure*}

In the $PS$ model, the comoving number density of collapsed dark matter haloes
of mass $M$ in the interval $dM$ is given by
\begin{equation}
\frac{dn}{dM} = -\sqrt{\frac{2}{\pi}} \frac{\delta_{c} \Pi}{\sigma(M,z)M}\frac{d\ln(\sigma \left(M,z\right))}{dM}Exp(-\frac{\delta_{c}^{2}}{2\left(\sigma(M,z)\right)^{2}}),
\end{equation}
where $\Pi$ is the comoving matter mean density of the Universe
and $\delta_{c}$ is the linearly extrapolated density threshold above
which structures collapse, i.e., $\delta_{c}=\delta_{L}(z=z_{col})$.
In an Einstein-de Sitter (EdS) model, an overdensity region collapses with a
linear contrast $\delta_{c}=1.686$ and is the value that we adopt for
our calculations. As a first approximation, we use this value in the EdS (we postponed for future work the use of spherical collapse model to calculate $\delta_{c}=\delta_{L}(z_{col})$). In principle, this election is not far from reality because for
homogeneus quintessence models coupled $ \delta_{c} $ does not separate too far from the EdS value (see \cite{nunes}).Recall that the $RCC$  model is equivalent to quintessence models \cite{ilso}.
Additionally, using the newtonian formalism for the spherical collapse, in appendix of the reference \cite{rcg}, 
were found a $\delta_{c}$ as function of the redshift, showing a slight deviation from the EdS value, being the best-fit $\delta_{c}=1.685$ very closed to our value used. 
The quantity $\sigma(M,z)=D(z)\sigma_{M}$ is the $rms$ linear fluctuation of density in spheres of radius $R$ containing a
mass $M$ and with growth factor $D(z)$. In our analysis, the $rms$
of the smoothed overdensity is given by
\begin{equation}
\sigma_{M} = \sigma_{8}(\frac{M}{M_{8}})^{-\gamma/3},
\end{equation}
where $M_{8}=6\times10^{14}\Omega_{m}h^{-1}M_{\odot}$, the mass
inside a sphere of radius $R_{8}= 8h^{-1}Mpc$, where $M_{\odot}$ is the solar mass. The index $\gamma$ is a function of the mass scale
and the shape parameter $\Gamma$ \cite{viana}

\begin{equation}
\gamma = (0.3\Gamma +0.2)[2.92+ \frac{1}{3}\log(\frac{M}{M_{8}})].
\end{equation}

We use $\Gamma =0.167$ \cite{sl}. We associate
galaxy clusters with dark matter haloes of the same mass. Our
analysis of the effects of a running cosmological constant on the number
of dark matter haloes is carried out by computing two quantities. The first quantity is the number
of haloes per unit of redshift in a given range of mass
\begin{equation}
\frac{dN}{dz} = \int_{4 \pi}{d\Omega}\int_{M_{inf}}^{M_{Sup}}{\frac{dn}{dM}\frac{dV}{dz d \Omega}dM}\,,
\end{equation}
where $\frac{dV}{dz d\Omega}$ is the comoving volume element, and is given by $r^{2}(z)/H(z)$,
where $r(z)=\int_{0}^{z}{dz/H(z)}$. In Figure 5,
we display the comoving volume element as a function of the redshift in
the $RCC$ model. We can see that there is a strong
dependence on the value of $\nu$. The $\Lambda CDM$ model is plotted
for comparison. In the right panel, we plot the comoving volume element compared
to the Einstein-de Sitter volume for all cases.

\begin{figure}[!h]
\begin{center}
\includegraphics[height= 7.5 cm,width=7.5cm]{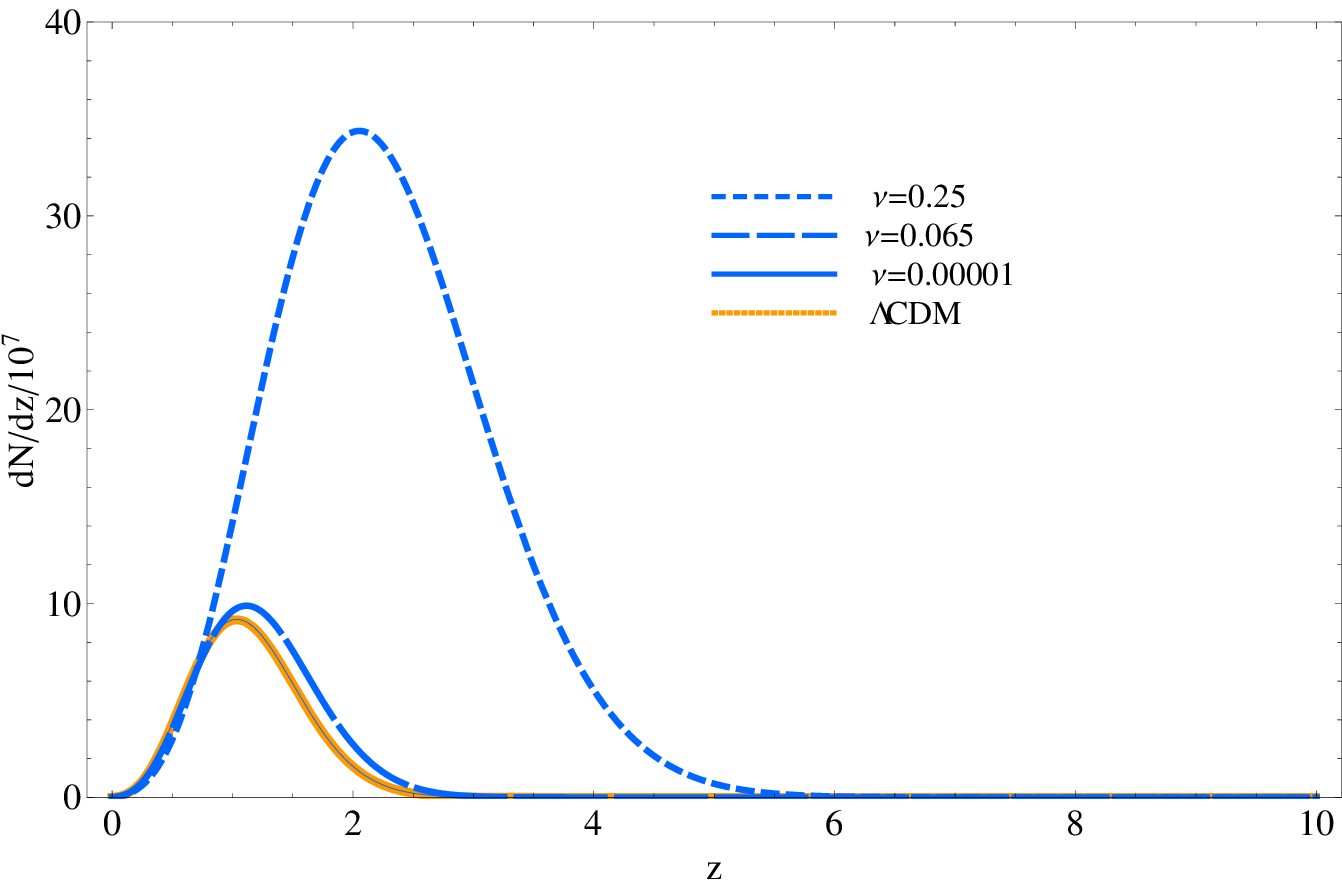}
\includegraphics[height= 7.5 cm,width=7.5cm]{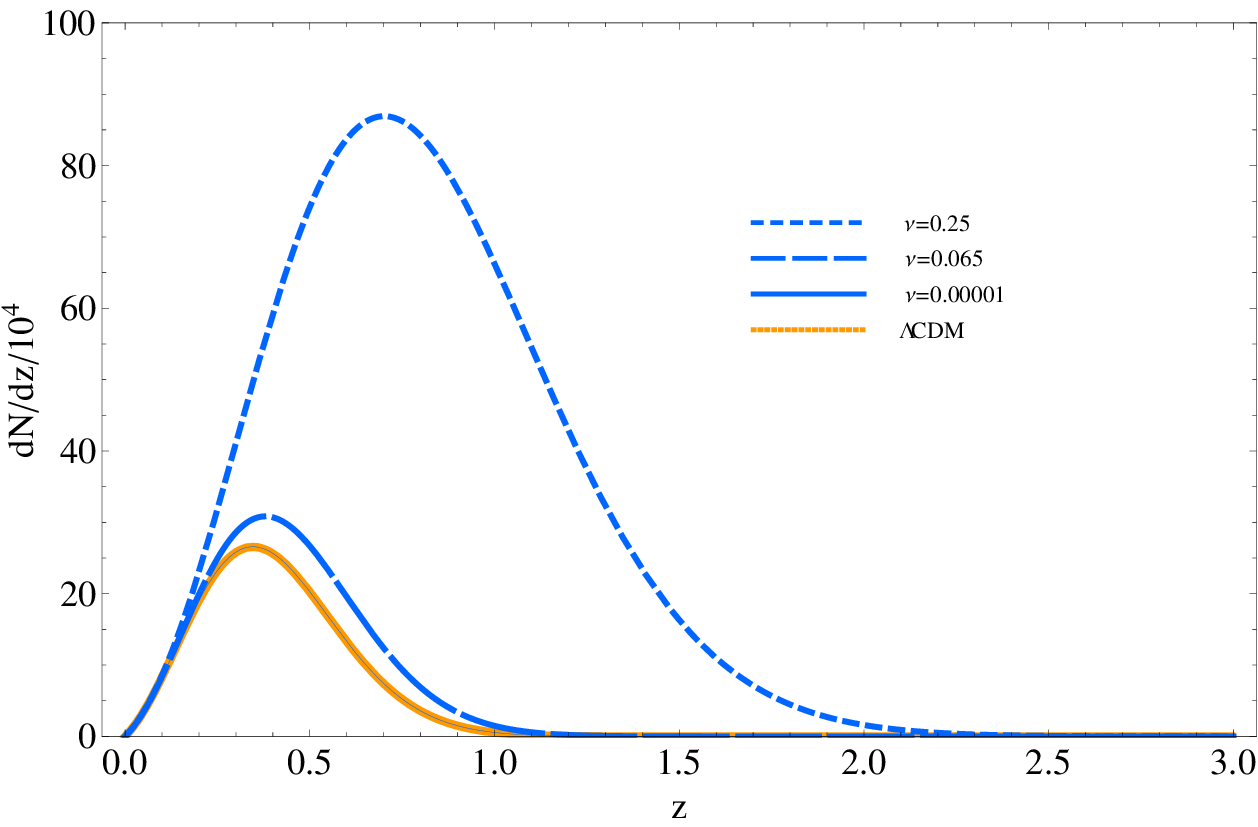}
\end{center}
\caption{The figure shows the evolution of the number counts with redshift
and the effect of the value of the parameter $\nu$ for
objects with a mass within the range $10^{13}<M/(h^{-1}M_{\odot})<10^{14}$.
The evolution of number counts for objects
with mass within the range $10^{14}<M/(h^{-1}M_{\odot})<10^{15}$ (below).
In both cases the curve $\nu=10^{-5}$ is indistinguishable from the $\Lambda CDM$ model.
We used a fixed value of $\Omega_{m0}=0.24$ for all curves.}
\label{fig6}
\end{figure}

\begin{figure}[htb]
\begin{center}
\includegraphics[height= 7.5cm,width=7.5cm]{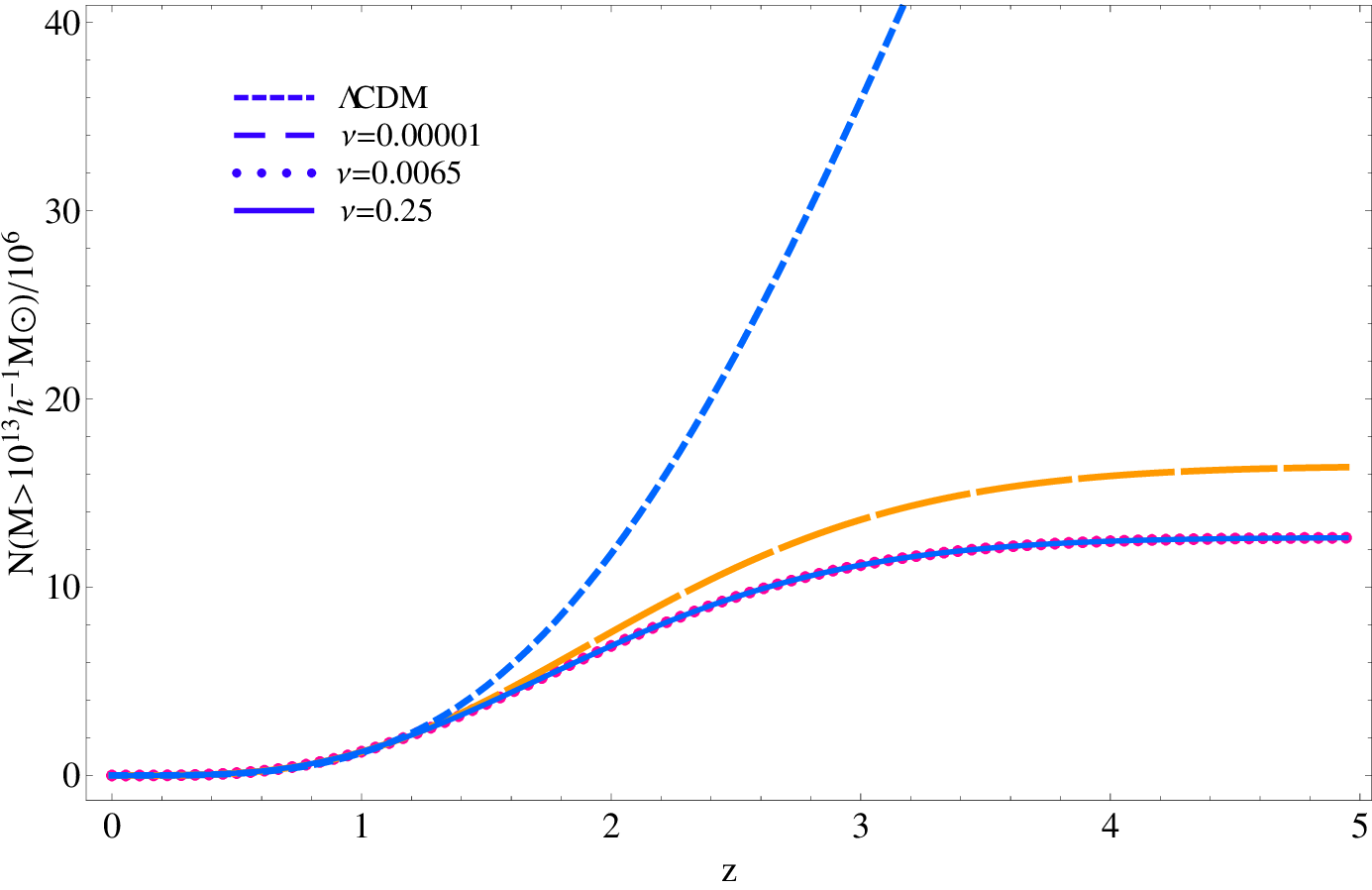}
\includegraphics[height= 7.5 cm,width=7.5cm]{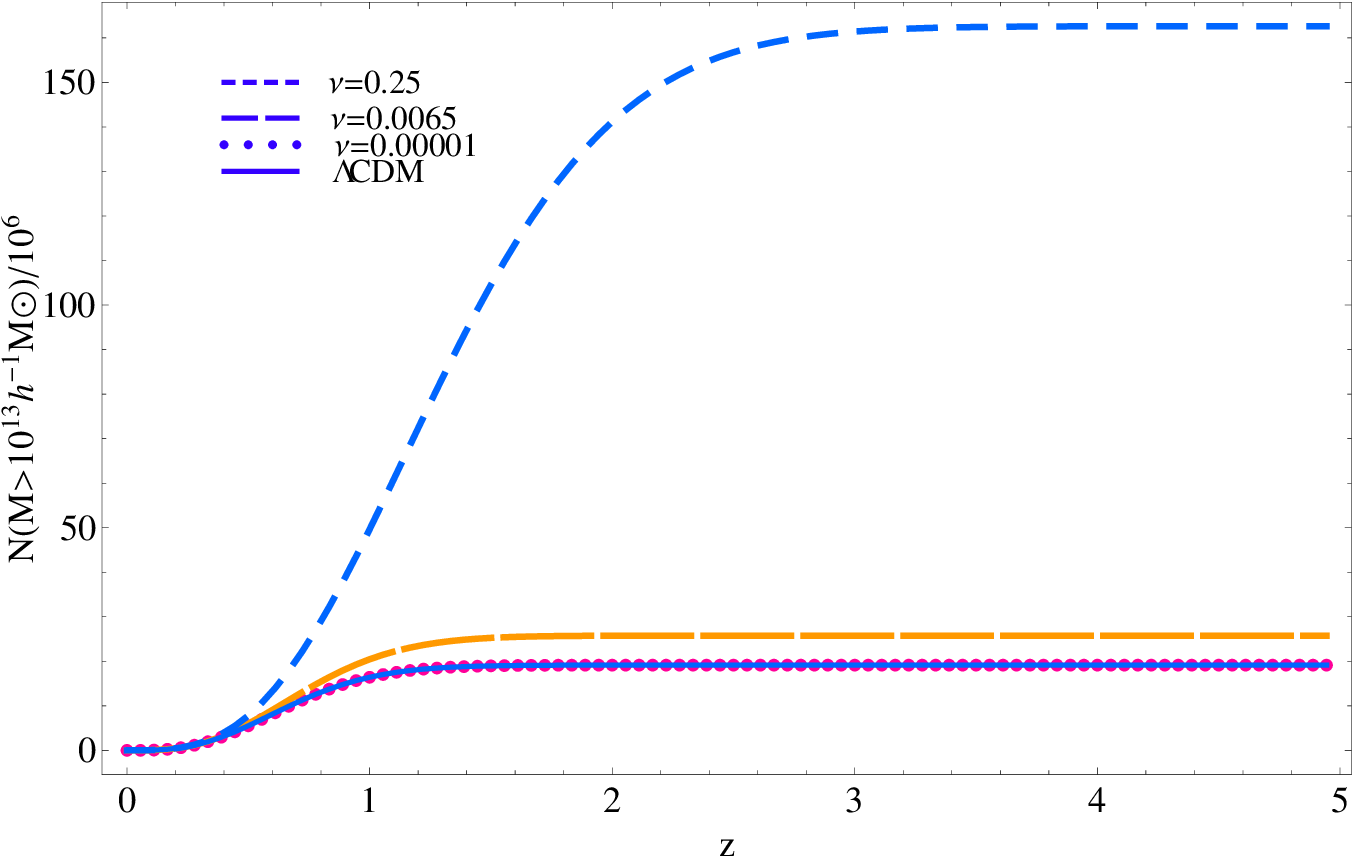}
\end{center}
\caption{The figures show the integrated number counts up to redshift $z=5$
for objects with mass $M>10^{13}h^{-1}M_{\odot}$ (top) and $M>10^{14}h^{-1}M_{\odot}$ (below).
}
\label{fig7}
\end{figure}

The other quantity that we compute is the all sky integrated number counts
above a given mass threshold, $M_{inf}$, and up to redshift $z$
\begin{equation}
N(z, M>M_{inf}) = \int_{4\pi} {d \Omega} \int_{M_{inf}}^{\infty}\int_{0}^{z}{\frac{dn}{dM}\frac{dV}{dz d \Omega}dz d \Omega}.
\end{equation}
This result is also called a cumulative mass function. To compute the
two quantities, we must choose a normalization of the number density
of haloes $n(M)$. This is commonly expressed in terms of $\sigma_{8}$. We
choose to normalize all models by fixing the number density of haloes
at redshift zero. At zero redshift, all models have the same
comoving background density $\Pi$ and growth factor $D$.
Our model fiducial is $\Lambda CDM$ ($\Omega_{m0}=0.24$, $h=0.72$) with $\sigma_{8}=0.9$ \cite{sl}.

Now, we can determine the dependence of these quantities on the
value of $\nu$. In Figure 6, we display the
number counts as a function of redshift. It is clear that
there is a strong dependence on the value assumed by the
parameter $\nu$. An increase in the value of $\nu$ produces an increase
in the value of $\frac{dN}{dz}$. Comparing the two panels,
we can see that there is a larger variation for greater values of mass.
The results for the total number of collapsed structures are displayed
in Figure 7. In the top panel, we show the integration in the range $M>10^{13}M_{\odot}$ and in the bottom
panel for $M>10^{14}M_{\odot}$; in both cases, $M_{sup}=10^{15}$. We do not use
strictly infinite, as $N(z,M>M_{inf})$ is dominated by the contribution
of the lower bound of the mass integration range.

The parameter $\nu$ can be considered a coupling parameter between dark energy (vacuum energy)
and dark matter. Models with more coupling have higher values of $dN/dz$.
This can be understood by the behavior of other observables. Equations
(37) and (38) have a dependences on
the growth factor $D(z)$, comoving energy density, and
the comoving volume element. An increase in the comoving volume
translates into an increase in the number counts; however,
average density and $D(z)$ decrease. Both effects produce the
observed results shown in Figures 6 and 7. This behavior is similar to models of
quintessence homogeneous \cite{nunes}.

\begin{figure*}[!h]
\begin{center}
\includegraphics[height= 6.7 cm,width=7.5cm]{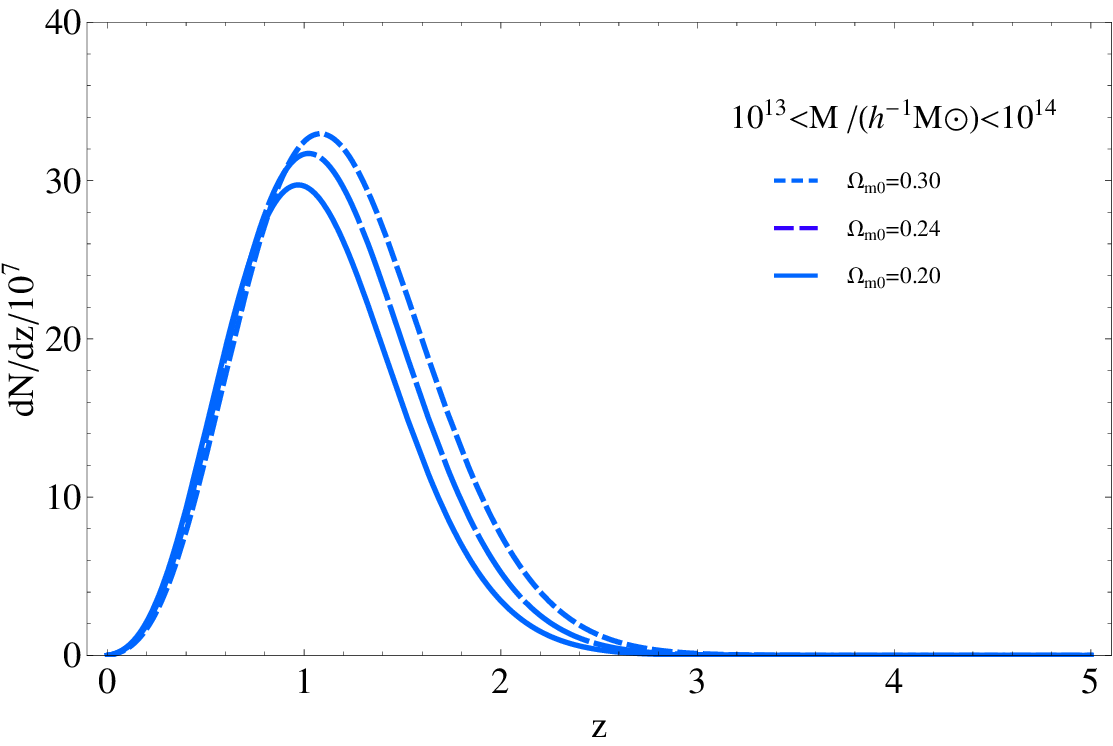}
\includegraphics[height= 6.5 cm,width=7.5cm]{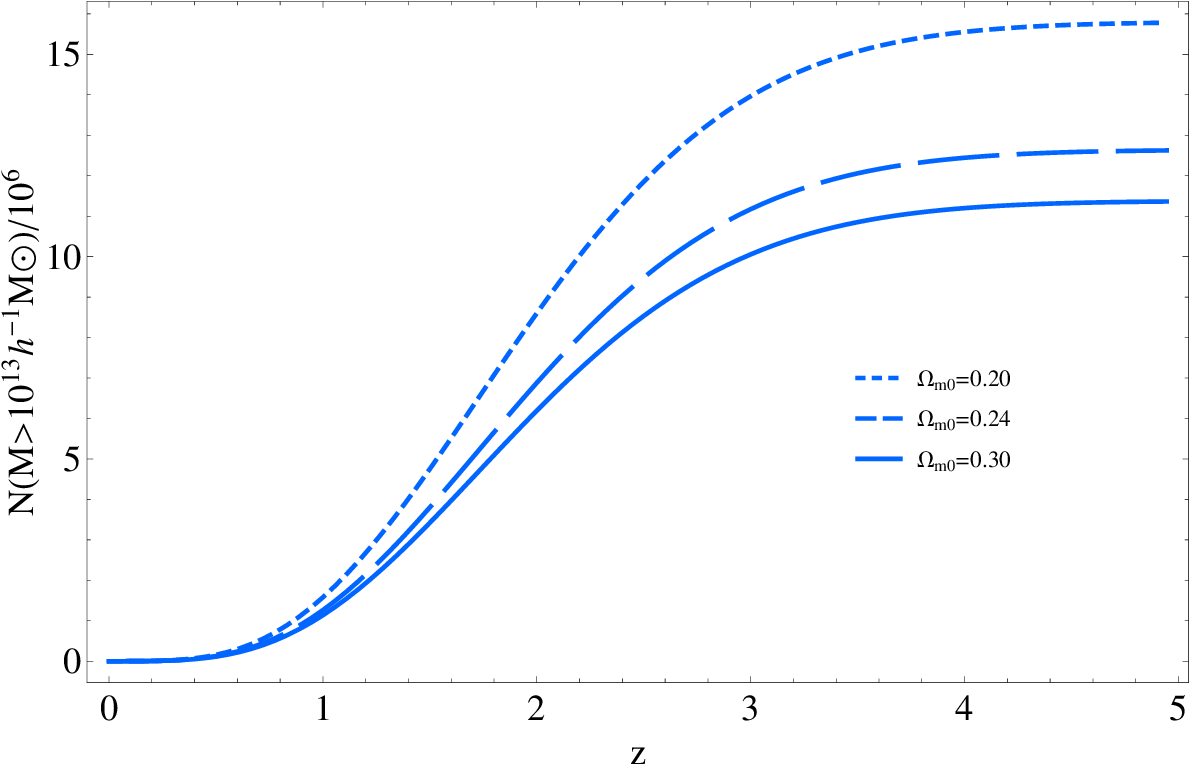}
\includegraphics[height= 6.5 cm,width=7.5cm]{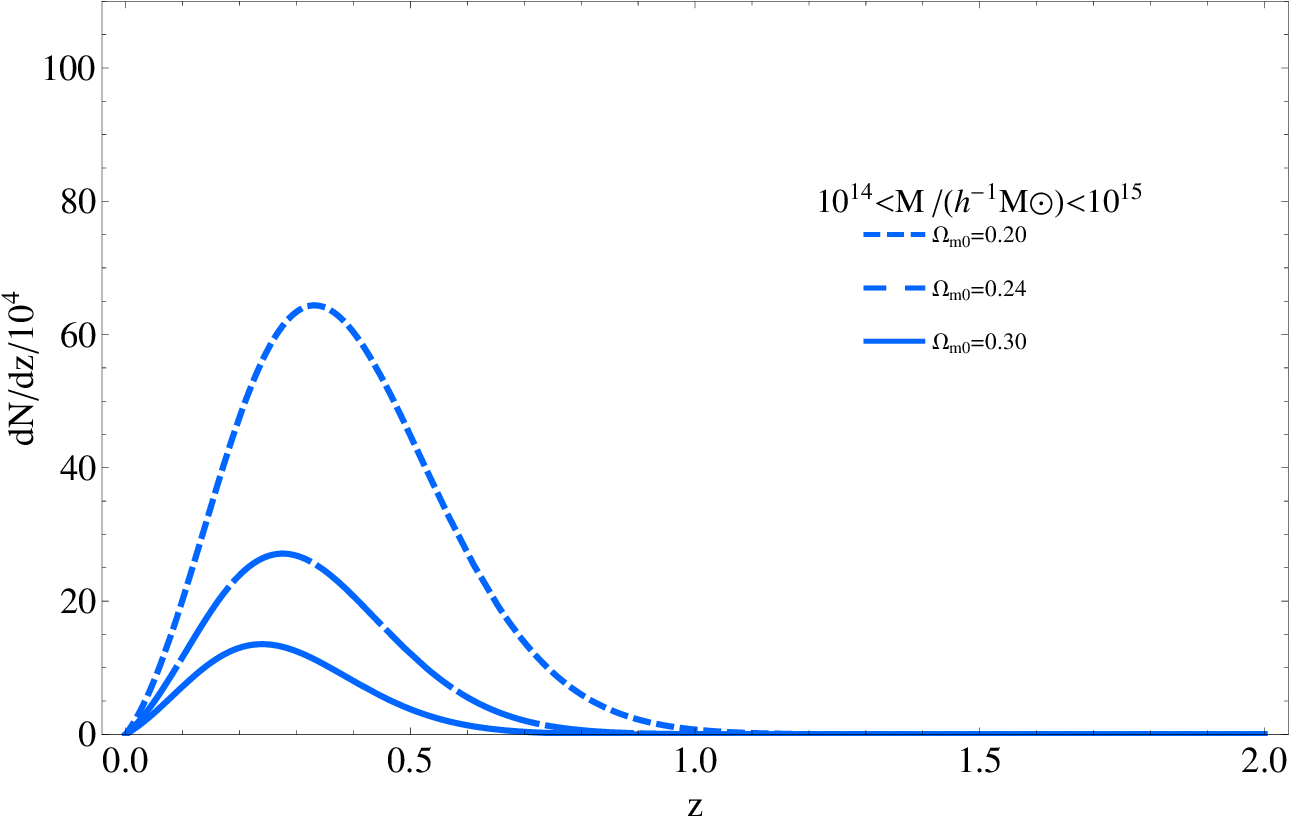}
\includegraphics[height= 6.5 cm,width=7.5cm]{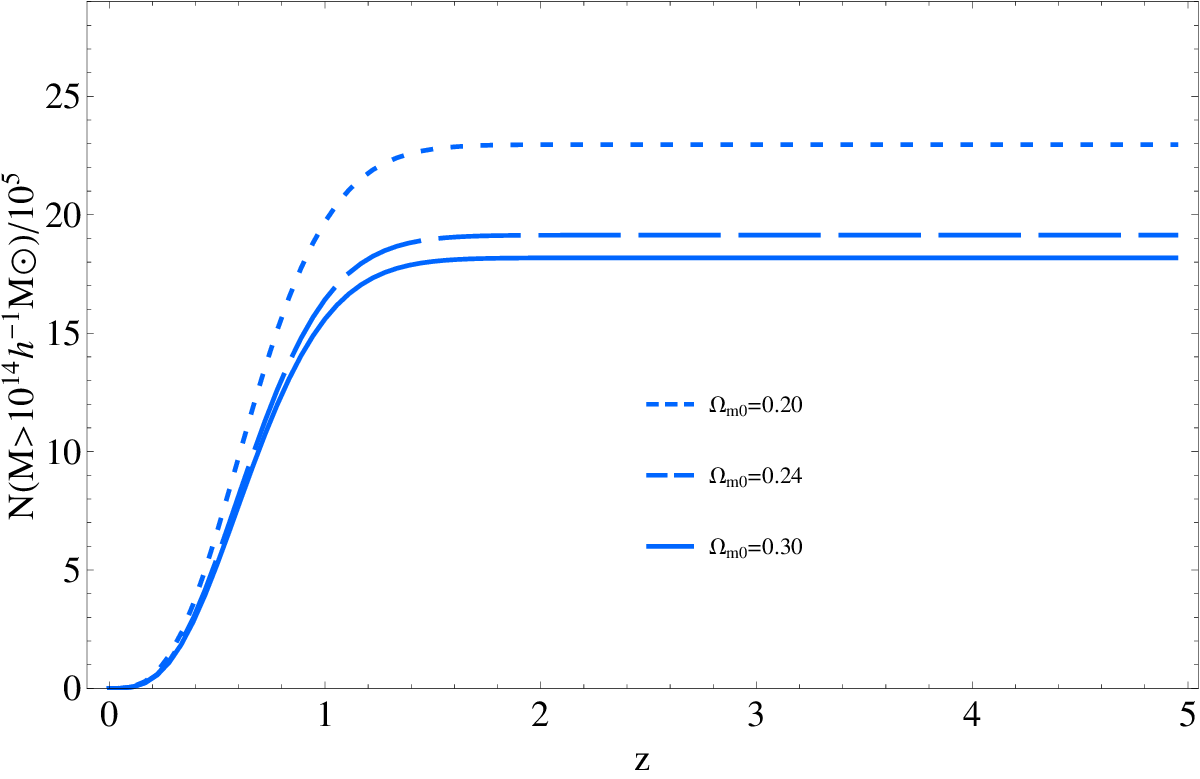}
\end{center}
\caption{The panels show the sensitivity on the parameter $\Omega_{m0}$. The right column shows the expected redshift distribution of $10^{13}<M/(h^{-1}M_{\odot})<10^{14}$ (upper) and $10^{14}<M/(h^{-1}M_{\odot})<10^{15}$ (lower) clusters.
The left column shows the integrated number counts of $M/(h^{-1}M_{\odot})< 10^{13}$(upper) and $M/(h^{-1}M_{\odot})< 10^{14}$(lower) clusters.}
\label{fig8}
\end{figure*}

\subsection{Cosmological Sensitivity on $\Omega_{m0}$ and $h$}

The cluster number counts depend on the cosmological parameters via the energy density,
growth factor, and comoving volume element. The cosmological dependence is implicit,
but is very strong. First, we consider the effects of changing $\Omega_{m0}$,
which are displayed in Figure 8. In the left column results
for $dN/dz$, with a mass between $10^{13}<M/(h^{-1}M_{\odot}) <10^{14}$ (top left)
and $10^{14}<M/(h^{-1}M_{\odot})<10^{15}$ (bottom left) are shown. The curves are for a flat
$RCC$ universe with $h=0.7$, $\nu=10^{-5}$ in all cases, and $\Omega_{m0}=0.20$ (dashed line),
$\Omega_{m0}=0.24$ (solid line) and $\Omega_{m0}=0.30$ (dotted line).
The right column shows the total number of clusters $N(z)$ for the same
values of all parameters. Several conclusions can be drawn from
Figure 8. Overall, a decrease in $\Omega_{m0}$ increases the
number of clusters at all redshifts (and vice versa).
The curve closest to $\Lambda CDM$ is the central curve.
Note that the dependence on $\Omega_{m0}$ is strong, for instance, a $16,7\%$
decrease in $\Omega_{m0}$ increases the total number of clusters $N(z)$
by $21\%$ for more massive structures (see the bottom right panel).

\begin{figure*}[!h]
\begin{center}
\includegraphics[height= 6.5 cm,width=7.5cm]{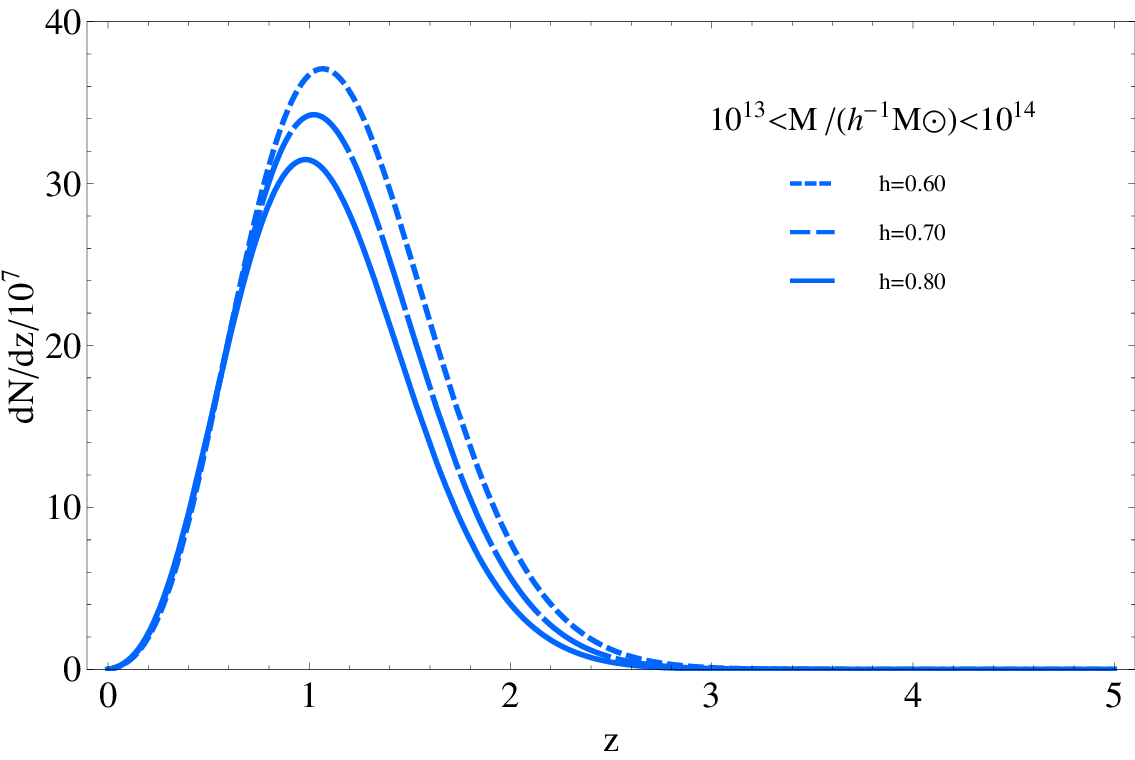}
\includegraphics[height= 6.5 cm,width=7.5cm]{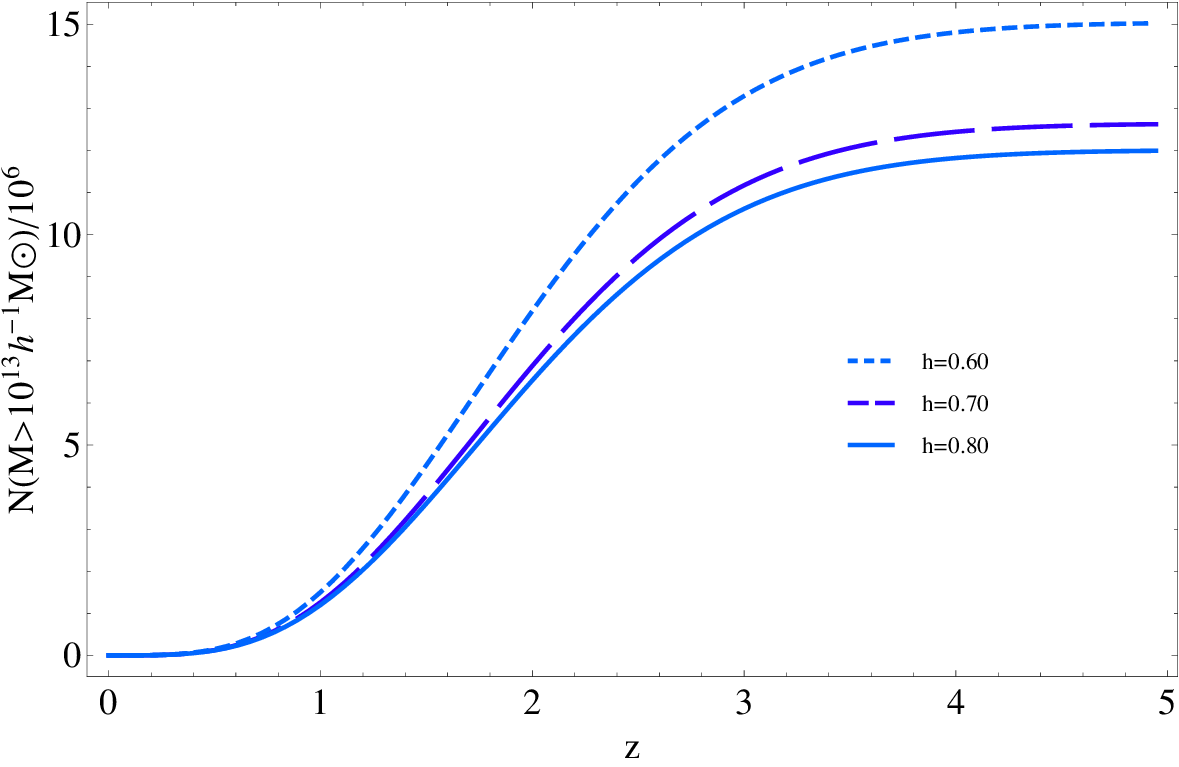}
\includegraphics[height= 6.5 cm,width=7.5cm]{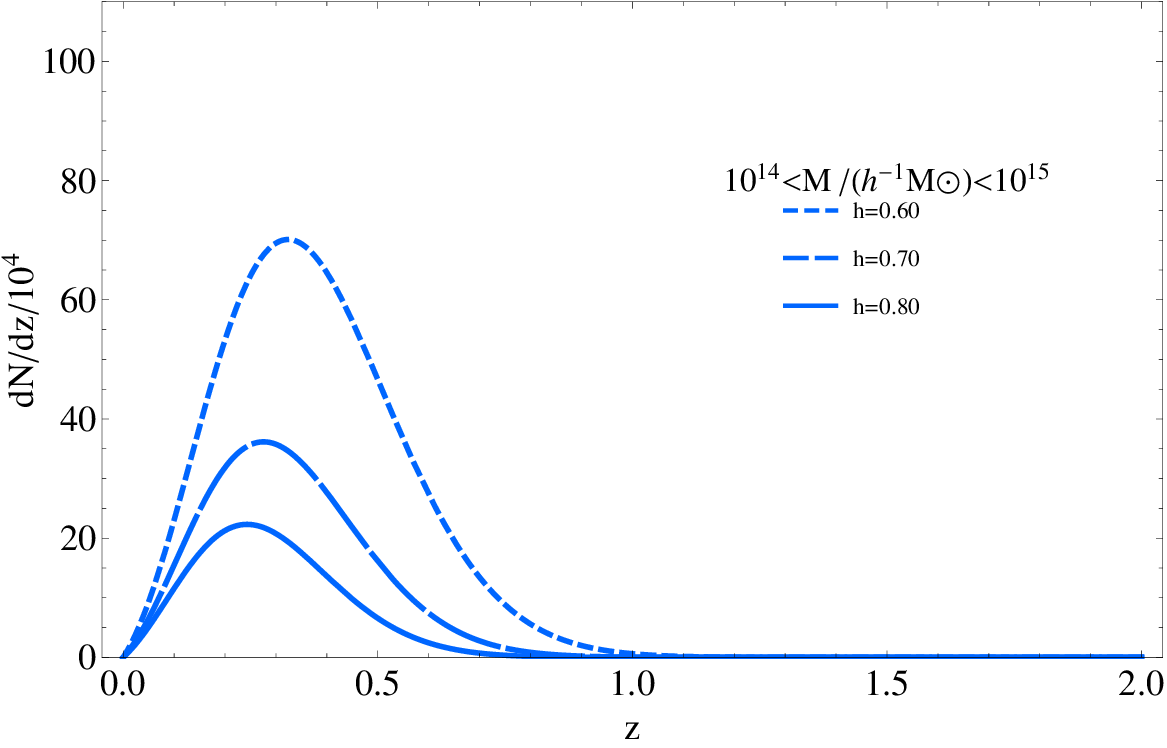}
\includegraphics[height= 6.5 cm,width=7.5cm]{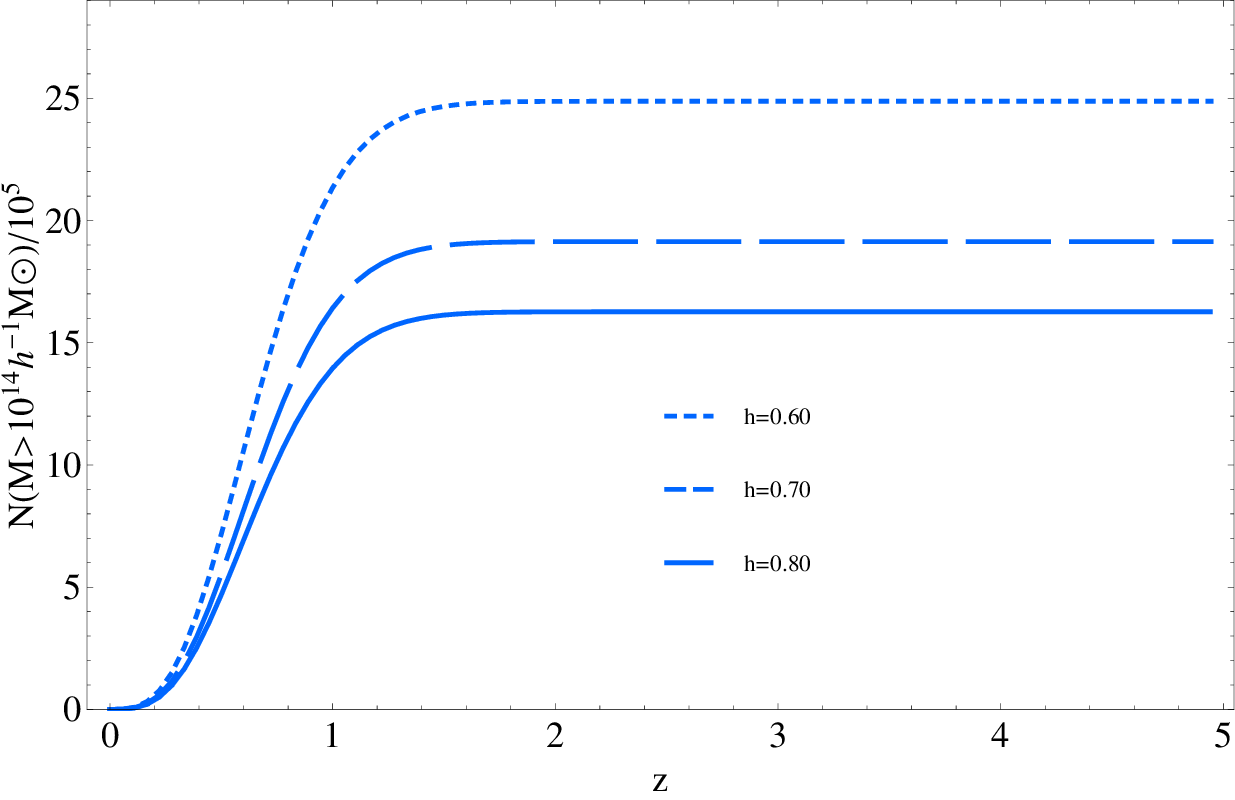}
\end{center}
\caption{The same as in Figure 8 for the sensitivity on the parameter $h$.}
\label{fig9}
\end{figure*}

Figure 9 demonstrates the effects of changing $h$.
Comparing Figure 8 to Figure 9, the quantitative behavior
of the observables ($N(z)$ and $dN/dz$) under changes in $h$ and $\Omega_{m0}$ are similar:
decreasing $h$ increases the total number of clusters, but does not
significantly change their redshift distribution for objects with
mass within the range of $10^{13}<M/(h^{-1}M_{\odot})<10^{14}$; however,
for masses between $10^{14}<M/(h^{-1}M_{\odot})<10^{15}$, the change
is greater.

\section{Summary and Discussions}

In this paper we have determined the cosmological implications
of the $RCC$ model using linear perturbations.
We used the longitudinal gauge to determine the density perturbations
of matter. In analyzing these perturbations we found, similar
to reference \cite{fabris}, that the perturbation of vacuum energy
density is proportional at the $\nu$ parameter (equation (25a)); thus,
when $\nu$ is zero, the standard scenario $\Lambda CDM$ is recovered.

Linear perturbation theory allows us to calculate the linear growth
rate $g(z)$ and compare our results both to those other models frequently studied
in the literature and to data for growth rate. In Figure 3, we have seen that for
values of $\nu \leq 10^{-4}$, our model predicts an overdensity similar to that of other models ($\Lambda CDM$,
$DGP$, and scalar-tensor theory). In Figure 3 (bottom),
we observed that for large values of $\nu$ the predicted $g(z)$ is far from the data.
These three data of
$g(z)$ are not sufficient to rule out a given model.
In Figure 4, we compared the matter power spectrum to the SDSS data
obtained by Percival et al. \cite{per} and
showed that a $RCC$ model with $\nu \approx 10^{-5}$ is compatible with $\Lambda CDM$.
The best-fit for our free parameter is $\nu=(1.29 \pm 2.39) \times 10^{-5}$.

We investigated the expected evolution of cluster number counts in
the $RCC$ cosmology.
In the present paper it has been studied using the $PS$ mass
function. We have shown that there is a
significant dependence on the cluster number counts in the $RCC$ model via the
amount of coupling between dark matter and vacuum energy. Increasing the
coupling, that is to say, increasing
the value of $\nu$ (see Figures 6 and 7), increase the cluster number counts.
This feature is compatible with the clustering properties in quintessence models \cite{mota}.

In general, our results for power spectrum and for linear growth rate
are compatible with the majority of other models. For example,
in the coupled quintessence model presented in reference \cite{pv},
the interaction must be very weak, on the order of $10^{-3}$,
for the model to be compatible with data from 2dFGRS.
Recently, in reference \cite{pelinson1} the perturbations
of a model called $\Lambda XCDM$ have been studied; this model includes two free parameters,
parameter $\nu$, which has the same meaning as in our case, and an equation
of state parameter $w_{X}$, the cosmon component. This models was proposed
and studied in reference \cite{lxcdm1,lxcdm2}, which showed that both components can interact. Reference \cite{pelinson2}
performed an analysis of the linear perturbation of these models;
their results were consistent with data from 2dFGRS.

In reference \cite{kim} a holographic model with infrared decay in CDM is considered. They use three
types of cut-off and,
in all case, the model has modes of growth for the density
contrast when the effective equation of state is between
$-1< w_{eff}<-1/3$. The authors have used a Newtonian
approximation, therefore, it is
only for the cut-off of the Hubble horizon that the model implies a
dark energy density proportional to the square of the Hubble
parameter (similar to our case).
We feel that it is necessary to have a fully relativistic approach
to the density contrast.

In reference \cite{carneiro}, a model with a cosmological
term that decays linearly with the Hubble parameter is considered. In that paper, the
authors consider a relativistic treatment of perturbations in the
synchronous gauge; the dark component is also perturbed. They
calculate the matter power spectrum and show that their results
are inconsistent with 2dFGRS data (the matter parameter
is very large $\Omega_{m} \approx  0.7$).

On the other hand, references \cite{nunes,nunes1}
explore the changes in the cluster number counts
predicted for models of homogeneous and inhomogeneous dark energy,
that is, two extreme limits for the evolution of dark energy in the
overdensity region. In the case of the homogeneous dark energy
density, the value of overdensity inside a given region is the same as in the background.
In the inhomogeneous dark energy case, there is a collapse with dark matter inside a given region.
The authors show that there is a deviation of up to $15\%$
with respect to $\Lambda CDM$ in the inhomogeneous case, and that
for both types of model, the largest deviations are observed
for massive structures $M>10^{16}$.
These results are consistent with our calculations.

Finally, it would be interesting to consider the other astrophysical implications
of the $RCC$ model with a modification of the spherical
collapse model, that is, the $\delta_{c}$ function of the redshift of collapse. Additionally, one could evaluate the profile of density contrast around the cluster,
and supercluster of void matter, using as a first
approximation, a $NFW$ (Navarro, Frenk, and While) profile \cite{NFW}.
Another important issue is the study of the concentration of haloes.
For example, following the prescription given in
reference \cite{navarro,navarro2}, we can investigate whether the
concentration of haloes in the $RCC$ model decreases with an
increase in the mass, as in the case of the $\Lambda CDM$ model.
This investigation would also determine the limits of the model.

\appendix

\section{Equations used to model DGP and Scalar-Tensor Theory}
In this appendix we write the equations used to determine the
growth factor in the DGP model and scalar-tensor theory.
The growth factor is defined in equation (28) and obeys equation \cite{uzan,leandros}
\begin{equation}
D''(k,a)+(\frac{3}{a}+\frac{H'(a)}{H(a)})D'(k,a)-\frac{3\Omega_{m0}}{2a^{5}H^{2}(a)}f(k,a)D(k,a)=0,
\end{equation}
with the initial condition $D(a) \approx a$ for $a \approx 0$
(in the matter-dominated era). This equation does not take into
account either perturbations in dark energy or anisotropic stress.
The function $f(k,a)$ expresses the connection between the metric
perturbations and the matter density perturbations. Therefore, the connection
depends on the particular gravity theory.
\subsection{DGP Model}
In the DGP model, in the case of flatness and matter being only found on the brane, $H(z)$ is given by \cite{dgp}
\begin{equation}
H^{2}_{DGP} = \sqrt{\Omega_{r_{c}}} + \sqrt{\Omega_{m0}a^{-3}+\Omega_{r_{c}}},
\end{equation}
where $\Omega_{r_{c}}=\frac{1}{4}(1-\Omega_{m0})^{2}$ and, in this
theory, $f(k,a)$ is given by \cite{uzan}
\begin{eqnarray}
f(k,a) &=&  \left(1+\frac{1}{3\beta}\right),
\end{eqnarray}
with
\begin{equation}
\beta = 1-\frac{H_{DGP}(a)}{H_{0}\sqrt{\Omega_{r_{c}}}}(1+\frac{aH'_{DGP}(a)}{3H_{DGP}(a)}).
\end{equation}
\subsection{Scalar-Tensor Theory}
Scalar-tensor theory is the simplest generalization of General Relativity
in which the fundamental constants are variable.
The function $f(k,a)$ in this case is given by \cite{uzan}
\begin{equation}
f(k,a) = \frac{G_{eff}(a)}{G_{eff}(a=1)}\left(1+\frac{1}{1+(k/ma)}\right),
\end{equation}
where $G_{eff}$ is the effective Newton constant, $a$ is the scale factor
and $a=1$ is the present value, and $m$ is the mass
of the scalar field $\Phi$ inducing a Yukawa cut-off
in the gravitational field. We used a simple ansatz given by \cite{leandros}
\begin{equation}
\frac{G_{eff}(a)}{G_{eff}(a=1)} = 1 + \xi(1-a)^{2}.
\end{equation}

In the case of the DGP model and scalar-tensor theory, the detection
of a value of $f(k,a) \neq  1$ would be a signature of
alternative theories of gravity. For the numerical calculations we
used $\xi = -0.2$.

\section*{Acknowledgement}
I wish to thank I. Shapiro for several discussions.
I also wish to thank W. Nesseris of the Niels Bohr International
Academy for sending some mathematica files. This research is a tribute to Luis Masperi.
The work was supported by the FAPEMIG (Minas Gerais, Brazil) agency and the Federal University of Juiz de Fora.

%
%

\end{document}